\pdfoutput=1

\documentclass[11pt,twoside,a4paper,cmspaper,final,collab]{cms-tdr}
\def\svnVersion{183694}\def\cmsCernNo{2013-007}\def\cmsCernDate{\today}\def\cmsMessage{Submitted to Physical Review Letters}

\begin{document}\cmsNoteHeader{EXO-11-016}

\hyphenation{had-ron-i-za-tion}
\hyphenation{cal-or-i-me-ter}
\hyphenation{de-vices}
\RCS$HeadURL: svn+ssh://svn.cern.ch/reps/tdr2/papers/EXO-11-016/trunk/EXO-11-016.tex $
\RCS$Id: EXO-11-016.tex 174952 2013-03-04 19:05:21Z alverson $
\providecommand{\coloron}{\ensuremath{\cmsSymbolFace{C}}\xspace}
\cmsNoteHeader{EXO-11-016}
\title{
Search for pair-produced dijet resonances in four-jet final states in \texorpdfstring{$\Pp\Pp$
collisions  at $\sqrt{s} = 7\TeV$}{pp collisions at sqrt(s) = 7 TeV}
}

\date{\today}

\abstract{
A search for the pair production of a heavy, narrow resonance
decaying into two jets
has been performed
using events collected in $\sqrt{s}=7\TeV$ $\Pp\Pp$ collisions with the CMS
detector at the LHC. The data sample corresponds to an integrated
luminosity of 5.0\fbinv.
Events are selected with at least four jets and two dijet combinations with similar dijet mass.
No resonances are found in the dijet mass spectrum.
The upper limit at 95\% confidence level on the product of the resonance pair
production cross section, the branching fractions into dijets, and the acceptance
varies from 0.22  to 0.005\unit{pb}, for resonance masses between 250  and 1200\GeV.
Pair-produced colorons decaying into
$\cPq\cPaq$  are excluded
for coloron masses between 250 and 740\GeV.
}

\hypersetup{%
pdfauthor={CMS Collaboration},%
pdftitle={Search for pair-produced dijet resonances in four-jet final states in pp collisions  at sqrt(s) = 7 TeV},
pdfsubject={CMS},%
pdfkeywords={CMS, physics, scalar gluon, dijet}}

\maketitle 

The high center-of-mass energy provided by the Large Hadron Collider (LHC) offers
opportunities to search for physics beyond the standard model (SM) and in particular
to search for new strongly interacting particles.
Searches for new resonances in the
dijet mass spectrum for the jets with the highest  transverse momentum ($\PT$)
have been performed at both the
Tevatron~\cite{Aaltonen:2008dn} and the
LHC~\cite{:2010bc,Khachatryan:2010jd,Chatrchyan:2011ns,Aad:2011fq,ATLAS:2012pu,CMS:2012yf}.
These searches were not optimized for pair-produced particles,
which are predicted by some models~\cite{Dobrescu:2007yp, Kilic:2008ub, Bai:2010dj}.
The ATLAS experiment has performed a search for pair production of
dijet resonances in four-jet events~\cite{Aad:2011yh,ATLAS:2012ds}
and excludes a scalar gluon  at 95\% confidence level (CL) in the mass
range between 100 and 287\GeV.

In this Letter, we report the results of a search for the pair production of a
narrow resonance, which decays into two jets, using the  dijet mass spectrum
in four-jet final states, measured with the Compact Muon Solenoid (CMS) detector
in $\Pp\Pp$ collisions  at the LHC at $\sqrt{s}=7\TeV$.
This search focuses on the high  mass range between 250 and 1200\GeV.
We define the common dijet mass as the average of the invariant masses of
the two dijets with the smallest difference in invariant mass, rejecting
events where the difference exceeds 15\% of the average value.
Models~\cite{Dobrescu:2007yp} predicting pair production through $\Pg\Pg$ interactions of color-octet vectors,
also called colorons (\coloron), and color-octet scalars ($S_8$) would give rise to such a signature.
In addition to $\cPq\cPaq$ decays, the coloron can also decay into a pair of color-octet scalars if the
mass of the coloron is greater than twice the mass of the two scalars.
Pair-produced colorons decaying to quark-antiquark
pairs (${\Pg\Pg} \rightarrow \coloron\coloron\rightarrow {\cPq\cPaq\cPq\cPaq}$)~\cite{Dobrescu:2007yp}
are used as the benchmark signal  in this analysis, although we separately consider
the possibility of decays to $S_8S_8$. As a third possibility, we consider an R-parity violating
SUSY model~\cite{Brust:2012uf,Evans:2012bf} in which pair-produced top squarks (stops) each decay
to $\cPq\cPaq$, as this leads to a similar final state.

The CMS detector is a multipurpose apparatus and is described in detail
in Ref.~\cite{cmsdet}.  The CMS coordinate system has its origin at
the center of the detector, the $z$ axis along the direction of the
counterclockwise
circulating proton beam, the $y$ axis normal to the LHC plane pointing vertically
upward, and the $x$ axis radially inward toward the center of the LHC ring.
We define $\phi$ to be the azimuthal angle, $\theta$ the polar angle,
and $\eta = -\ln[\tan(\theta/2)]$  the pseudorapidity.  The
central feature of the CMS apparatus is a superconducting solenoid
with a 6\unit{m} internal diameter, operating at a central field strength of
3.8\unit{T}.  Within the field volume are, in  order of increasing radius, a silicon pixel and strip
tracker, a high-granularity PbWO$_4$ crystal electromagnetic calorimeter (ECAL),
and a brass/scintillator hadron calorimeter (HCAL). All three systems have
both barrel and endcap components, with $|\eta|=$ 2.5 (3.0) the outermost extent
of the tracker (calorimeters). The ECAL and HCAL cells are grouped into towers,
projecting radially outward from the origin.
Outside of the field volume an iron/quartz-fiber
hadron calorimeter covers the forward region ($3<|\eta|<5$).
A  muon system encloses the central and endcap calorimeters out to  $|\eta|$ = $2.4$.

The data sample used for this analysis was
collected in 2011 and corresponds to an integrated luminosity of  $5.0$~fb$^{-1}$.
The triggers used for the analysis require the presence of at least four jets,
based on information from the calorimeters. Each jet must have  $|\eta|$ less than 3.0 and
 $\PT$ greater than 70 or 80\GeV, depending on the running period.
This trigger is 99.5\% efficient for events
with four leading (highest $\PT$) jets, each with a transverse momentum exceeding 110\GeV.

For offline reconstruction, we employ the CMS particle-flow algorithm~\cite{pflow} in
the region $|\eta| \le 2.5$ to reconstruct objects used in jet determinations.
This algorithm uses calorimeter information and reconstructed tracks to identify
electrons, muons, photons, and both neutral and charged hadrons.
Jets are reconstructed from particle-flow objects using the
anti-\kt algorithm with a distance parameter
0.5~\cite{ak5,Cacciari:2005hq}.  Jet energy  is corrected to account for
the nonlinearities and nonuniformities in the response
of the calorimeters, as
determined from Monte Carlo (MC) simulation, test beam, and collision
data~\cite{Chatrchyan:2011ds}.
Additional corrections accounting for the effect of multiple $\Pp\Pp$ collisions
per bunch crossing are also applied~\cite{Cacciari:2007fd, Cacciari:2011ma}.

We require events to have at least one good primary vertex with a $z$
position within 24\unit{cm} of the center of the detector and with a transverse distance
from the beam spot of less than 2\unit{cm}.  A set of jet
quality criteria are applied to remove possible instrumental and
non-collision backgrounds~\cite{Chatrchyan:2011tn}.
All data as well as all simulated signal
events passing these selection criteria also satisfy standard jet
identification requirements~\cite{JME-10-003-PAS}.
We require that events have at least
four jets, each with $\PT>110$\GeV and $|\eta|<2.5$.
We require the two jets in each possible pair to have a
separation $\Delta{R_{jj}}=\sqrt{(\Delta\eta)^2+(\Delta\phi)^2}\ge0.7$.
This ensures a
negligible overlap between the jets.
We calculate the dijet-mass combinations from the four leading jets and choose the one with the smallest
$\Delta m /m_\text{avg}$, where $\Delta m$ is the mass
difference between the two dijets and $m_\text{avg}$ is their average mass.
We require  $\Delta m  <0.15 m_\text{avg}$, which is approximately three times the
dijet mass resolution of 4.5\%.

The benchmark signal events are simulated using the \MADGRAPH v5~\cite{Alwall:2011uj}
event generator with the CTEQ6L1 parton distribution functions (PDF)~\cite{Lai:2010nw},
and \PYTHIA v6.4.26~\cite{Sjostrand:2006za}  parton showering  and
hadronization. The generated events are further processed through
a \GEANTfour~\cite{Agostinelli:2002hh} simulation of the CMS detector.
The assumed width of the simulated coloron resonance is negligible compared with the
experimental resolution.  The dominant background arises from QCD processes
resulting in four or more jets.  Studies of this background are performed
using a sample of simulated QCD events
also generated using \MADGRAPH.

\begin{figure}[htb]
  \begin{center}
   \includegraphics[width=\linewidth]{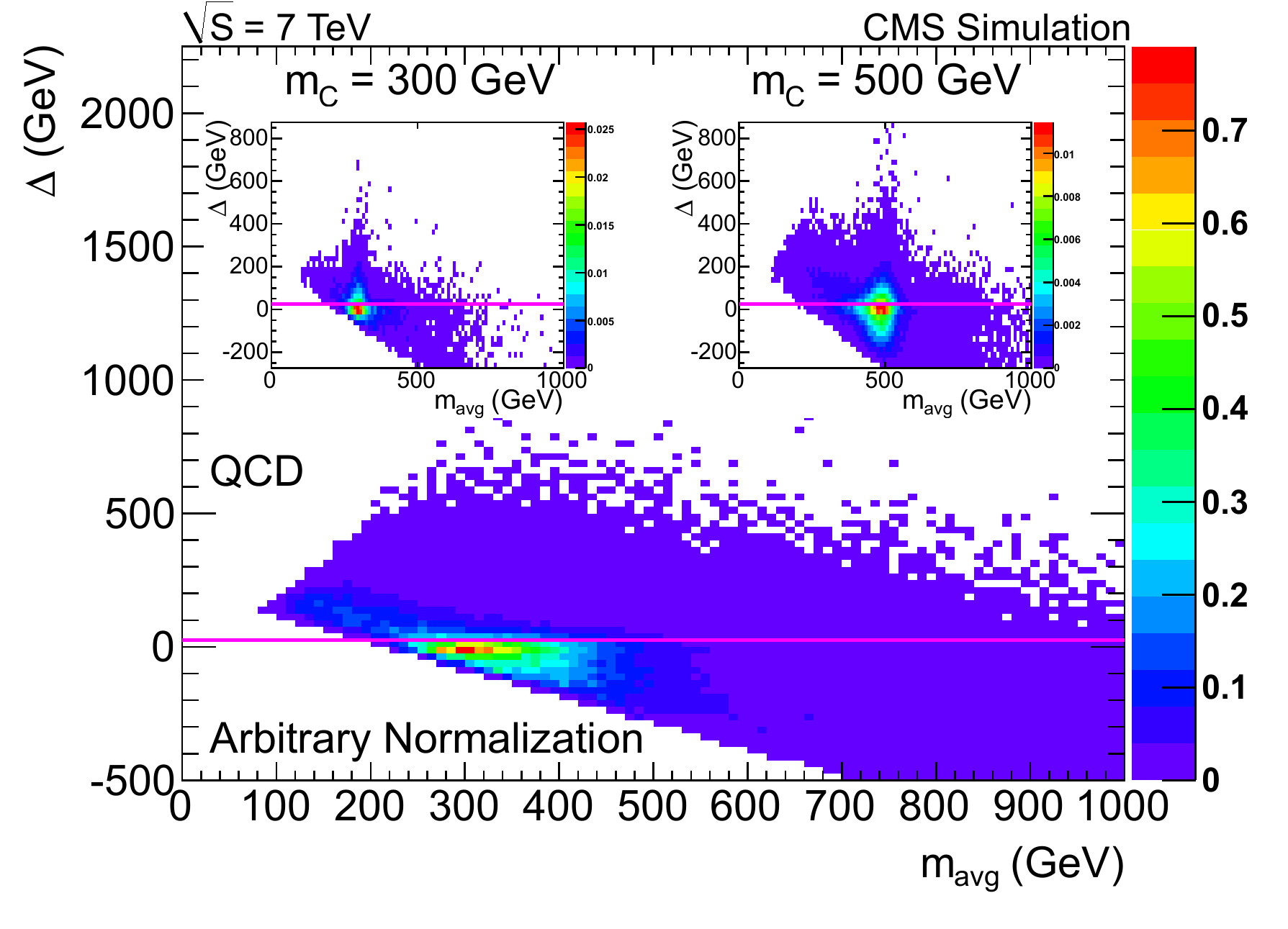}
    \caption{
The distribution  of  $\Delta= \sum_{i=1,2} (\PT)_i  - m_\text{avg}$ versus $m_\text{avg}$ for
QCD multijet background alone (main plot) and background plus coloron signal (insets), for simulated events.
We remove events with an entry below the horizontal line at 25\GeV.
There are two entries per event (one for each of
the two dijet pairs).
    }
    \label{fig:diagdemo}
  \end{center}
\end{figure}

For each dijet we define a quantity $\Delta$ as the difference
between the scalar sum of the transverse momenta of the two jets in the dijet and the average
pair mass in the event: $ \Delta= \sum_{i=1,2} (\PT)_i - m_\text{avg}.$
Figure~\ref{fig:diagdemo} shows the distribution of $\Delta$ versus
$m_\text{avg}$ for simulated QCD background events as well as for coloron signal events.
Due to the selection requirements, we observe a broad structure at around
$m_\text{avg}= 300\GeV$ from QCD events~\cite{Chivukula:1991zk}.
To remove this structure, thus leaving a smoothly falling dijet mass spectrum, we require
$\Delta > 25\GeV$ for each of the two dijets in the event.
This requirement reduces the QCD background by more than an order of magnitude while retaining
approximately 25\% of the signal.

\begin{figure}[htb]
  \begin{center}
   \includegraphics[width=\linewidth]{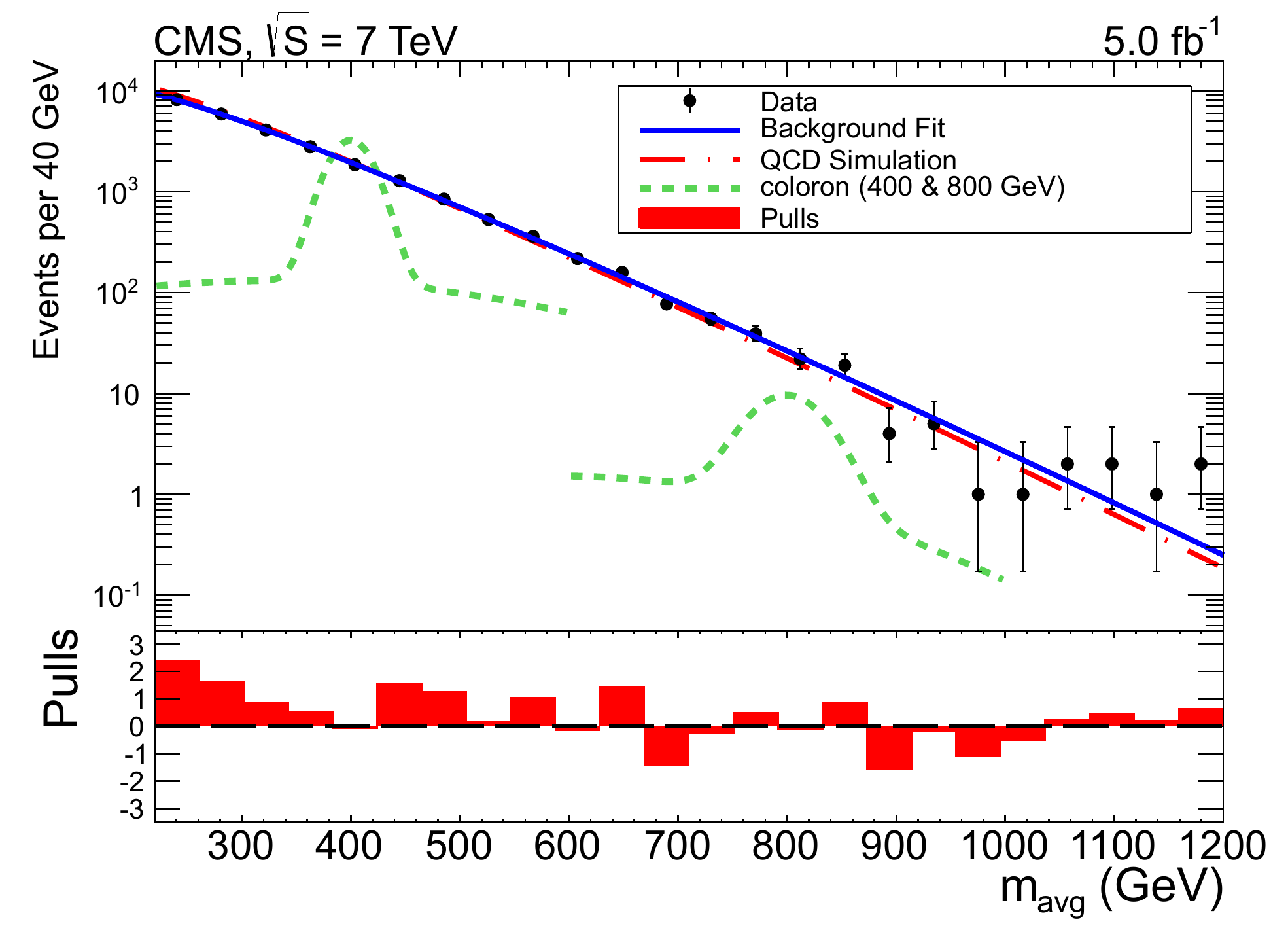}
    \caption{The  average paired dijet mass distribution (black points) in data
compared with a smooth background fit  using Eq.~(\ref{eqParam}) (blue solid curve) and the result of the fit of
the same function (dash-dotted red curve) to QCD simulated data (not shown).
Also shown are examples of simulation of hypothetical coloron resonances decaying 100\% to
$\cPq\cPaq$ (dashed green curves) with masses $m_\coloron = 400$ and 800\GeV.
The bin-by-bin fit pulls  are shown below.
}
    \label{fig:datanullFit}
  \end{center}
\end{figure}

Figure~\ref{fig:datanullFit} shows the paired dijet mass spectrum in
data with all the selection criteria applied.
The observed mass spectrum extends up to 1200\GeV.
We obtain a prediction for
the QCD background by fitting the data to a smooth parametrization:
\begin{equation}
\frac{\rd\sigma}{\rd{}m_\text{avg}} =
\frac{P_{0} (1 - m_\text{avg}/\sqrt{s})^{P_{1}}}{(m_\text{avg}/\sqrt{s})^{P_{2} + P_{3} \ln
(m_\text{avg}/\sqrt{s})}},
\label{eqParam}
\end{equation}
\noindent where $P_0$, $P_1$, $P_2$, and $P_3$ are free parameters.
This functional form has been used in previous searches for dijet
resonances~\cite{Chatrchyan:2011ns}.
The fit to the data and the normalized QCD simulation are given in
Fig.~\ref{fig:datanullFit} by solid and dashed-dotted curves, respectively.
The fit has a $\chi^2/\text{d.o.f}$ of 0.94 over the full $m_\text{avg}$ mass range.
Although there is an apparent bias toward positive pull values in the low mass region, such a
bias would result in the quoted limits being conservative in this region.

The signal shapes from colorons and stops have negligible difference and we use a single parameterization for both.
It is modeled by the sum of two separate Gaussian functions:
one Gaussian describes the core and the other
the tails, with widths and normalizations determined from a fit to simulated signal
events at each assumed mass value.  The dijet mass
resolution described by the rms of the core Gaussian is approximately 4.5\%, the
dijet mass tail described by the rms  of the other Gaussian  is between 150 and 250\GeV,
and the fraction of the core Gaussian varies from 85\% at 300\GeV to 45\% at 1000\GeV.
The signal  acceptance, listed in Table~\ref{myacc},
varies from 0.4\% for a coloron with mass 200\GeV to 12.1\% for a
coloron with mass 1000\GeV.
The acceptance for the stop signal is larger than that for the coloron signal because
the stop production  model includes $\cPq\cPaq$ interactions and has a different final
state angular distribution.

We search for pair production of dijet resonances by fitting
the background parameterization in Eq.~(\ref{eqParam}) plus a
signal hypothesis at an assumed mass to  the data.
The signal magnitude is a free parameter in the fit
for each fixed value of signal mass.
We restrict the fit to dijet masses above 220\GeV in order to avoid the
threshold due to the jet $\PT$ requirement.
With this requirement the dijet mass spectrum is described by both the simulation
and the background parameterization.
The maximum value of the likelihood obtained from
a background-only fit is denoted by
$L_0$ and the likelihood from a background plus signal fit by $L_S$.
The local significance is defined as $\sqrt{-2\ln(L_0/L_S)}$.
The results of this analysis are shown in Fig.~\ref{fig:datanullFit}.
We find that the largest fluctuation in the pair-produced dijet mass spectrum occurs
for a hypothetical resonance mass of 1130\GeV and has a local
statistical significance of $2.6\sigma$.  The global significance is reduced
to  $1.2\sigma$ after taking into account the trials factor~\cite{lookelse}
within the full mass range of this search.
We conclude that there is no  evidence for pair-produced
narrow dijet resonances in the data.

Upper limits are placed on the product of the production cross section of the pair-produced
resonances, the square of the branching fraction to dijets, and the detector acceptance. The dijet mass
for the limit is required to be above 250\GeV
to ensure a full coverage of the low mass tail of the resonance between 220 and 250\GeV.
To set upper limits we use a modified-frequentist
method (CL$_\mathrm{s}$)~\cite{Read:2002hq,Junk:1999kv}.
We fit the signal+background hypothesis to the data, allowing both the signal strength and the
background parameters free to vary.
The sources of systematic uncertainties consist of a luminosity uncertainty
of 2.2\%~\cite{CMS:2012rua} and a signal acceptance uncertainty of 10\%.
The latter is determined by the jet energy scale
uncertainty (2.2\%) and the jet energy resolution uncertainty (10\%)~\cite{Chatrchyan:2011ds}.
The variation in expected signal yields due to PDF uncertainties is negligible.
The uncertainties in the
luminosity, the signal acceptance due to jet energy scale and resolution, and the parameters of the
background function are all treated as nuisance parameters and
expressed as log-normal distributions with their central values and
uncertainties.  The observed and expected limits are calculated using the CL$_\mathrm{s}$ method
with a one-sided profile likelihood test statistic.

\begin{figure}[hbt]
  \begin{center}
     \includegraphics[width=\linewidth]{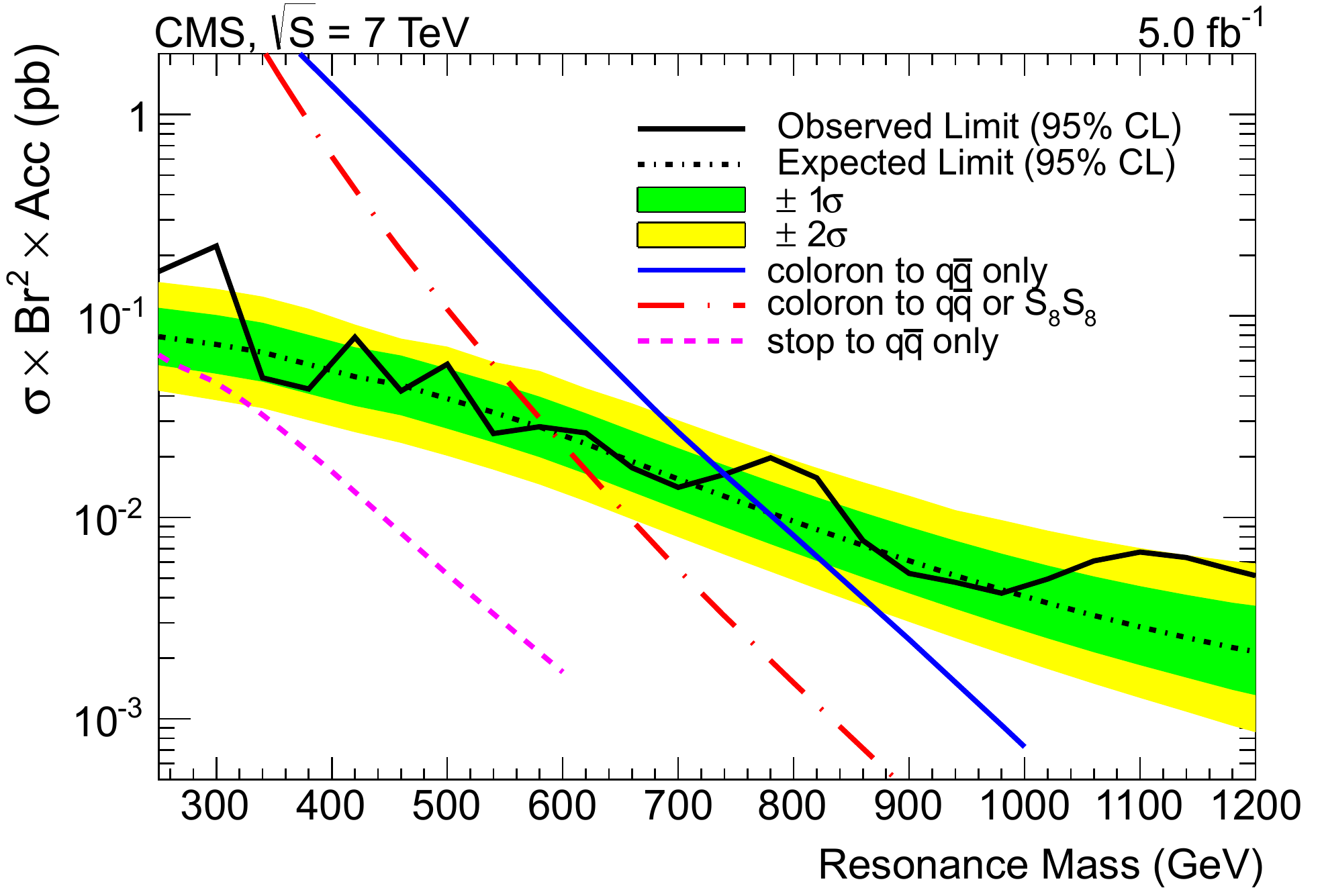}
     \caption{
The observed and expected 95\% CL limits
on the product of the resonance pair
production cross section, the square of the branching fraction to dijets, and the detector
acceptance, given
by the solid and dot-dashed black curves, respectively.
The shaded regions indicate the $1\sigma$ and $2\sigma$ bands around the expected limits.
Predictions of a coloron model and a SUSY model are also shown.
}
    \label{fig:limit}
  \end{center}
\end{figure}

Figure~\ref{fig:limit} shows the observed and expected 95\% CL limits, the $1\sigma$ and
$2\sigma$ uncertainty bands around the expected limits, and predictions
from the coloron and SUSY models. The observed limit on the product of
the resonance pair production cross section, the branching fractions
into dijets, and the acceptance varies from 0.22 to 0.005\unit{pb} for
resonance masses between 250 and 1200\GeV.
The limits are generally applicable for
pair-produced resonances, each decaying to dijets, and they
are compared with calculations
for the coloron model~\cite{Dobrescu:2007yp} described above.
At 95\% CL we exclude  pair production of colorons with mass m$_C$ in the range
$250< m_\coloron < 740\GeV$,  assuming that colorons have flavor-universal couplings
and decay only  into $\cPq\cPaq$~\cite{Bai:2010dj}.
Assuming the branching fraction of colorons into $\cPq\cPaq$ is reduced due to
competition with a $\coloron \rightarrow S_8 S_8$ channel where  m$_{S_8}$ = 150\GeV
and $\tan\theta = 0.3$ (the suppression factor of gluon coupling to  $\cPq\cPaq$
compared with the analogous QCD coupling)~\cite{Bai:2010dj},
we exclude pair production in the range  $250< m_\coloron < 580\GeV$.
This analysis is not sensitive to the pair-produced $S_8$, where the color-octet scalars
decay exclusively to $\cPq\cPaq$.
We also compare the results with those of a SUSY model for pair-produced stops, where  the stops decay
exclusively to $\cPq\cPaq$ and R-parity is
violated~\cite{Brust:2012uf,Evans:2012bf}. The calculation is done at
next-to-leading order (NLO) with next-to-NLO
corrections~\cite{Beenakker:2011fu,Beenakker:2009ha,Beenakker:1996ch,Kulesza:2008jb,Kulesza:2009kq}.

\begin{table*}[hpbt]
\topcaption{
The acceptances for the coloron and stop models after applying all selection criteria.
Most of the variation in the acceptance as a function of resonance mass is due to the jet \pt requirement.
}
\label{myacc}
\centering
\begin{scotch}{c|ccccccccc}
mass [\GeVns{}] & 200 & 300 &400 &500 &600 &700 &800 &900 &1000 \\ \hline
coloron acceptance & 0.4\% & 2.2\% & 5.2\% & 8.0\% & 9.6\% & 10.6\% & 11.6\% & 11.8\% & 12.1\%   \\
stop acceptance & 0.9\% & 3.6\% & 7.9\% & 10.7\% & 12.9\% &---  &---  & --- &---    \\
\end{scotch}
\end{table*}

In summary, a search for pair production of a narrow dijet resonance has been performed
with the CMS detector using 5.0\fbinv of
$\sqrt{s}$ = 7\TeV $\Pp\Pp$ collisions produced at the LHC.
The paired dijet mass spectrum is found to be a smooth  distribution and is
in agreement with the predictions of the standard model.
Upper limits are reported on the product of the
production cross section, the branching fractions into dijets, and the acceptance
of a  pair-produced dijet resonance having a width
negligible compared with the experimental resolution.
At 95\% CL, the pair production of colorons is excluded
for coloron masses between
250 and 740\GeV assuming that a coloron decays  100\%
into $\cPq\cPaq$, or between 250 and 580\GeV assuming  that coloron
decays into $\cPq\cPaq$ compete with decays into $S_8 S_8$.
The search significantly extends previous results~\cite{ATLAS:2012ds}.

We congratulate our colleagues in the CERN accelerator departments for the
excellent performance of the LHC machine. We thank the technical and administrative
staff at CERN and other CMS institutes, and acknowledge support from BMWF and
FWF (Austria); FNRS and FWO (Belgium); CNPq, CAPES, FAPERJ, and FAPESP (Brazil);
MES (Bulgaria); CERN; CAS, MoST, and NSFC (China); COLCIENCIAS (Colombia); MSES (Croatia);
RPF (Cyprus); MoER, SF0690030s09 and ERDF (Estonia); Academy of Finland, MEC, and
HIP (Finland); CEA and CNRS/IN2P3 (France); BMBF, DFG, and HGF (Germany); GSRT (Greece);
OTKA and NKTH (Hungary); DAE and DST (India); IPM (Iran); SFI (Ireland); INFN (Italy);
NRF and WCU (Korea); LAS (Lithuania); CINVESTAV, CONACYT, SEP, and UASLP-FAI (Mexico);
MSI (New Zealand); PAEC (Pakistan); MSHE and NSC (Poland); FCT (Portugal);
JINR (Armenia, Belarus, Georgia, Ukraine, Uzbekistan); MON, RosAtom, RAS and RFBR (Russia);
MSTD (Serbia); SEIDI and CPAN (Spain); Swiss Funding Agencies (Switzerland); NSC (Taipei);
ThEP, IPST and NECTEC (Thailand); TUBITAK and TAEK (Turkey); NASU (Ukraine); STFC (United
Kingdom); DOE and NSF (USA).

\bibliography{auto_generated}

\cleardoublepage \appendix\section{The CMS Collaboration \label{app:collab}}\begin{sloppypar}\hyphenpenalty=5000\widowpenalty=500\clubpenalty=5000\textbf{Yerevan Physics Institute,  Yerevan,  Armenia}\\*[0pt]
S.~Chatrchyan, V.~Khachatryan, A.M.~Sirunyan, A.~Tumasyan
\vskip\cmsinstskip
\textbf{Institut f\"{u}r Hochenergiephysik der OeAW,  Wien,  Austria}\\*[0pt]
W.~Adam, E.~Aguilo, T.~Bergauer, M.~Dragicevic, J.~Er\"{o}, C.~Fabjan\cmsAuthorMark{1}, M.~Friedl, R.~Fr\"{u}hwirth\cmsAuthorMark{1}, V.M.~Ghete, N.~H\"{o}rmann, J.~Hrubec, M.~Jeitler\cmsAuthorMark{1}, W.~Kiesenhofer, V.~Kn\"{u}nz, M.~Krammer\cmsAuthorMark{1}, I.~Kr\"{a}tschmer, D.~Liko, I.~Mikulec, M.~Pernicka$^{\textrm{\dag}}$, D.~Rabady\cmsAuthorMark{2}, B.~Rahbaran, C.~Rohringer, H.~Rohringer, R.~Sch\"{o}fbeck, J.~Strauss, A.~Taurok, W.~Waltenberger, C.-E.~Wulz\cmsAuthorMark{1}
% ...
\vskip\cmsinstskip
\dag:~Deceased\\
1:~~Also at Vienna University of Technology, Vienna, Austria\\
2:~~Also at CERN, European Organization for Nuclear Research, Geneva, Switzerland\\
% ...

\end{sloppypar}
\end{document}